\documentclass[12pt,a4paper]{article}
\usepackage{epsf}

\begin{document}
\section{Introduction}

The ability to prepare arrays with made to measure electronic 
properties has generated much recent activity\cite{r1}. 
This was 
made possible by the development of synthetic methods for the 
preparation of, so called, quantum dots. For our purpose these dots 
act like designer atoms. The properties of individual dots can be 
designed by the choice of composition and size. Size is a particular 
variable we wish to emphasize. Quantum dots are aggregates of tens or 
hundreds of atoms (or molecules) and so are nanoscale building 
blocks. The size of the dots is however small enough so that, since 
the highest lying electron(s) are confined to the volume of the dot, 
the electronic energy levels are still discrete. For a chemist, it is 
convenient to think of the higher-most electrons of each dot as the 
valence electron of an atom\cite{r2}.

These dots can self assemble into a 
planar array. For, e.g., Ag nanodots, the packing is hexagonal.
The reason why adjacent dots in such an array may have an unequal 
charge distribution is related to their nano-scale size. Unlike the 
smaller ordinary atoms, a quantum dot does not require a high energy 
cost to induce an electron transfer from a neighboring dot. This is 
technically stated by saying that the charging energy of a dot is 
atypically low. Electrostatic considerations suggest that the 
charging energy scales inversely with size and this has been 
experimentally born out. Metallic quantum dots have charging energy 
of well below an eV while
atoms have charging energies of at least 
several eV's.

The current attention on architecture with quantum dots building 
blocks is partly motivated by potential applications of such 
nanoscale quantum devices. In this paper we explore how corresponding 
computations can be carried out. The technical problem is that the 
Coulombic repulsion between two electrons (of opposite spins) that 
occupy the same dot cannot be described in a one electron 
approximation. It requires allowing for correlation of electrons. The 
simplest model Hamiltonian that can describe the Coulombic (or 
charging energy) effects is due to Hubbard\cite{1}. This paper is a 
study of exploring a computationally tractable method for solving for 
the states of the Hubbard Hamiltonian for a lattice. The development 
makes essential use of the nature of the intended applications. At 
low temperatures, only those states of the array that are quite close 
to the ground state are important\cite{r3}.

The Hubbard model is the simplest model one can use to study
the Mott metal-insulator transition, i.e. the transition from a metallic to an
insulating phase driven by
correlations between electrons\cite{r4}.  
Despite the extensive efforts in studying the Hubbard model \cite{1,2,3}
there are still difficulties in extending the one-dimensional results
to higher-dimensions.
The difficulties mainly come from the electron correlations inherent
in this model, which becomes more  complicated in two and three-dimensions.
Several methods have been developed
to treat these  problems such
as exact diagonalization methods\cite{4}, quantum Monte-Carlo 
simulations\cite{5}
and a variety of approximation techniques like mean-field theory\cite{6},
Green's function decoupling schemes\cite{7}, functional integral formulations
\cite{8}, variational approach\cite{9} and perturbation expansion \cite{10}.
It is already known that the numerical methods suffer from the finite-size
effect and perturbation calculations are not useful when dealing with the
intermediate-to-strong correlations. In this context, using the real-space
renormalization group method seem to be a promising direction because of
its non-perturbative nature. In fact, this method
has been applied to quantum lattice systems\cite{11}. For
example, the density matrix renormalization group\cite{12} has
provided a revolutionary way to obtain reliable results for the 
one-dimensional Hubbard
model and the numerical renormalization group for interacting
finite Fermi systems\cite{12a}
and for excitations in atoms\cite{12b}.
To apply the renormalization group method  to higher dimensions, the 
simplest way
is to use the block renormalization group method(BRG)\cite{13}. The 
BRG method has
already been applied to interacting electrons in two-dimensional systems
for both bipartite lattices\cite{14} and more recently for nonpartite
lattices such as the triangular one\cite{15}.

In this paper, we present the real-space block renormalization group 
equations for fermion
systems on a triangular lattice with hexagonal blocks.
By using operator transformations, we obtain the
conditions that make the renormalized Hamiltonian in the bare Hubbard model
have the same structure as the old one, i.e. the renormalization group
equations in the
parameters space with constant dimension will remain the same, 
without any proliferation of
couplings. Finally, using numerical calculations we
show that the metal-insulator transition occurs at the critical value 
of $U/t \approx 12.5$.
\section{Real-space renormalization group}

In this section, we present a general computational procedure for applying
the BRG method for interacting fermions on a lattice using the Hubbard
Hamiltonian. As an example we will take the hexagonal block on a triangular
lattice. The Hubbard Hamiltonian can be written as follows,

\begin{equation}
H=-t\sum_{<i,j>,\sigma }[c_{i\sigma }^{+}c_{j\sigma
}+H.c.]+U\sum_in_{i\uparrow }n_{i\downarrow }-\mu \sum_i(n_{i\uparrow
}+n_{i\downarrow }),
\end{equation}

where $t$ is the nearest-neighbor hopping term, $U$ is the local repulsive
interaction and $\mu $ is the chemical potential. $c_{i\sigma
}^{+}(c_{i\sigma })$ creates(annihilates) an electron with spin $\sigma $ in
a Wannier orbital located at site $i$; the corresponding number operator is 
$n_{i\sigma }=c_{i\sigma }^{+}c_{i\sigma }$ and $<$ $>$ denotes the
nearest-neighbor pairs. H.c. denotes the Hermitian conjugate.

The essence of the BRG method is to map the above many-particle Hamiltonian
on a lattice to
a new one with fewer degrees of freedom and with the same low-lying 
energy levels
\cite{16}. Then the mapping is repeated leading to a final Hamiltonian for
which an exact solution can be obtained. The procedure can be divided 
into three steps:
First divide the $N-$site lattice into appropriate $n_s-$site blocks labeled by
$p(p=1,2,...,N/n_s)$ and separate the Hamiltonian $H$ into intrablock part
$H_B$ and interblock $H_{IB}$

\begin{equation}
H =H_B+H_{IB}= \sum_pH_p+\sum_{<p,p^{\prime }>}V_{p,p^{\prime }},
\end{equation}

where

\begin{equation}
H_p = -t\sum_{<i^{(p)},j^{(p)}>} [c_{i^{(p)}\sigma }^{+}c_{j^{(p)}\sigma
}+H.C.]+U\sum_{i^{(p)}}n_{i^{(p)}\uparrow }n_{i^{(p)}\downarrow }-\mu
\sum_{i^{(p)}}(n_{i^{(p)}\uparrow }+n_{i^{(p)}\downarrow }),
\end{equation}

and

\begin{equation}
V_{p,p^{\prime }} =-t\sum_{<i^{(p,b)},j^{(p^{\prime },b)}>}
[c_{i^{(p,b)}\sigma }^{+}c_{j^{(p^{\prime },b)}\sigma }+H.C.]
\end{equation}

in which $i^{(p)}$ denotes the $ith$ site on the $pth$ block and $i^{(p,b)}$
denotes the border site of the block $p$.

The second step is to
solve $H_p$ exactly to get the eigenvalues $E_{p_i}$ and eigenfunctions
$\Phi _{p_i}(i=1,2,...,4^{n_s}).$ Then we can build the eigenfunctions of
$H_B $ by direct multiplication of $\Phi _{p_i}$, which can be written as
$|\Psi _B(i_{1,}i_2,...,i_{N/n_s})>=|\Phi _{1i_1}>|\Phi _{2i_2}>...|\Phi
_{N/n_s}i_{N/n_s}>(i_1,i_2,...\in \left\{ 1,2,...,4^{n_s}\right\} )$.

The last step is to
treat each block as one site on a new lattice and the correlations between
blocks as hopping interactions. The original Hilbert space has four 
states per site.
By following the above procedure one obtains an equivalent Hamiltonian with
  $(4^{n_s})^{N/n_s}=4^N,$ degrees of freedom, which is the same as the original
Hamiltonian. But in the realistic case, if we only care about the 
properties related
to some special energy levels of the system it is not necessary  to 
keep all the
states for a block to obtain the new Hamiltonian. For example, when studying
the metal-insulator-transition\cite{17}, we may only need to consider the
ground state and the first excited state energies.

The above scheme is a general procedure for applying the BRG method. In
order to make the new Hamiltonian more tractable, it is desirable to make it
have the same structure as the original one, i.e. the reduction in 
size should not be accompanied by a proliferation of new
couplings. Then we can use the iteration procedures to solve the model. To
achieve this goal it is necessary to keep only 4 states in step 2, which can
be understood from the following renormalized intrasite Hamiltonian. The
four selected states are taken to be

\begin{eqnarray}
|\Phi _{p1}>\equiv & |0>_p^{\prime }, \\
|\Phi _{p2}>\equiv &c_{p\uparrow \downarrow }^{\prime +}|0>_p^{\prime
}=|\uparrow \downarrow >_p^{\prime }, \\
|\Phi _{p3}> \equiv &  c_{p\uparrow }^{\prime +}|0>_p^{\prime }=|\uparrow
>_p^{\prime }, \\
|\Phi _{p4}> \equiv & c_{p\downarrow }^{\prime +}|0>_p^{\prime }=|\downarrow
>_p^{\prime },
\end{eqnarray}

where $c_{p\sigma }^{\prime +}(c_{p\sigma }^{\prime })$ is the
creation(annihilation)operator of the block state $|\sigma >_p^{\prime }$
and their corresponding energies are $E_i$ $(i=1,2,3,4)$.

\medskip Our next task is to rewrite the old Hamiltonian $H=H_B+H_{IB}$ in
the space spanned by the truncated basis

\begin{equation}
H^{\prime }=\sum_{\Psi _B^{Truncated} \stackrel{\_}{\Psi }_B^{Truncated}}
|\Psi _B^{Truncated}><\Psi _B^{Truncated}|H|\stackrel{\_}{\Psi }%
_B^{Truncated}><\stackrel{\_}{\Psi }_B^{Truncated}|,
\end{equation}

where the truncated basis is given by
\begin{equation}
|\Psi_B^{Truncated}(i_{1,}i_2,...,i_{N/n_s})>=|\Phi _{1i_1}>|\Phi
_{2i_2}>...|\Phi _{N/n_s}i_{N/n_s}>(i_1,i_2,...\in \{1,2,3,4\})
\end{equation}

In order to avoid proliferation of additional couplings in $H^{\prime },$
the four states kept from the block cannot be arbitrarily chosen. Some
definite conditions must be satisfied in order to make $H^{\prime }$ have
the same structure as $H$. Substituting $H$ into $H^{\prime }$ and using the
product of different operators (see Table I)  we can  get the expression
for $H_p$

\begin{eqnarray}
H_p &=&|0>_p^{\prime }E_1<0|_p^{\prime }+|\uparrow \downarrow >_p^{\prime
}E_2<\downarrow \uparrow |_p^{\prime }+|\downarrow >_p^{\prime
}E_4<\downarrow |_p^{\prime }+|\uparrow >_p^{\prime }E_3<\uparrow
|_p^{\prime }  \nonumber \\
&=&E_1+(E_3-E_1)n_{p,\uparrow }^{\prime }+(E_4-E_1)n_{p,\downarrow }^{\prime
}+(E_1+E_2-E_3-E_{4)}n_{p,\uparrow }^{\prime }n_{p,\downarrow }^{\prime }
\end{eqnarray}
Note that by keeping only four states from the block states in the beginning
gives no other extra couplings in the new Hamiltonian.

\begin{center}
Table I: The product of different operator transformations$^{*}$
\begin{eqnarray*}
\begin{tabular}{|c|c|c|c|c|}
\hline
& $<0|^{\prime }$ & $<\uparrow |^{\prime }$ & $<\downarrow |^{\prime }$ & $%
<\uparrow \downarrow |^{\prime }$ \\ \hline
$|0>^{\prime }$ & $1-n_{\uparrow }^{\prime }-n_{\downarrow }^{\prime
}+n_{\uparrow }^{\prime }n_{\downarrow }^{\prime }$ & $c_{\uparrow }^{\prime
}-n_{\downarrow }^{\prime }c_{\uparrow }^{\prime }$ & $c_{\downarrow
}^{\prime }-n_{\uparrow }^{\prime }c_{\downarrow }^{\prime }$ & $%
c_{\downarrow }^{\prime }c_{\uparrow }^{\prime }$ \\ \hline
$|\uparrow >^{\prime }$ & $c_{\uparrow }^{\prime +}-c_{\uparrow }^{\prime
+}n_{\downarrow }^{\prime }$ & $n_{\uparrow }^{\prime }-n_{\uparrow
}^{\prime }n_{\downarrow }^{\prime }$ & $c_{\uparrow }^{\prime
+}c_{\downarrow }^{\prime }$ & $-n_{\uparrow }^{\prime }c_{\downarrow
}^{\prime }$ \\ \hline
$|\downarrow >^{\prime }$ & $c_{\downarrow }^{\prime +}-c_{\downarrow
}^{\prime +}n_{\uparrow }^{\prime }$ & $c_{\downarrow }^{\prime
+}c_{\uparrow }^{\prime }$ & $n_{\downarrow }^{\prime }-n_{\downarrow
}^{\prime }n_{\uparrow }^{\prime }$ & $n_{\downarrow }^{\prime }c_{\uparrow
}^{\prime }$ \\ \hline
$|\uparrow \downarrow >^{\prime }$ & $c_{\uparrow }^{\prime +}c_{\downarrow
}^{\prime +}$ & $-n_{\uparrow }^{\prime }c_{\downarrow }^{\prime +}$ & $%
c_{\uparrow }^{\prime +}n_{\downarrow }^{\prime }$ & $n_{\uparrow }^{\prime
}n_{\downarrow }^{\prime }$ \\ \hline
\end{tabular}
\end{eqnarray*}

$^{*}$In this table the product reads $|0>^{\prime }<0|^{\prime }=1-n_{\uparrow
}^{\prime }-n_{\downarrow }^{\prime }+n_{\uparrow }^{\prime }n_{\downarrow
}^{\prime } $, etc.
\end{center}

\medskip Comparing the above intrasite Hamiltonian with Eq.(1), we get the
next conditions in order to copy the intrasite structure of the old
Hamiltonian, i.e. $E_3=E_4.$ Because of the additional vacuum
energy $E_1$ in the new Hamiltonian we rewrite
the intrasite part of Eq.(1) as

\begin{equation}
H_B=U\sum_in_{i\uparrow }n_{i\downarrow }-\mu \sum_i(n_{i\uparrow
}+n_{i\downarrow })+K\sum_iI_i
\end{equation}

where we introduce another parameter $K$ to the original system and $I_i$ is a
unit operator. The new intrasite Hamiltonian is given by

\begin{equation}
H_B^{\prime }=(E_1+E_2-2E_3)\sum_pn_{p\uparrow }^{\prime }n_{p\downarrow
}^{\prime }-(E_1-E_3)\sum_p(n_{p\uparrow }^{\prime }+n_{p\downarrow
}^{\prime })+E_1\sum_pI_p.
\end{equation}
Then the renormalized parameters $U,\mu$ and $ K$ can be obtained from
the following relations

\begin{eqnarray}
U^{\prime } &=&E_1+E_2-2E_3 \\
\mu ^{\prime } &=&E_1-E_3 \\
K^{\prime } &=&E_1
\end{eqnarray}
in which $E_1,E_2$ and $E_3$ are functions of the old parameters $t,U,\mu, K$.

For the half-filled case, $\mu =U/2,$ $H_B$ can be
expressed as

\begin{equation}
H_B=U\sum_i(\frac 12-n_{i\uparrow })(\frac 12-n_{i\downarrow })+K\sum_iI_i
\end{equation}
with the initial value of $K=-\frac U4$.
By using the particle-hole symmetry,  $E_1=E_2$, the renormalization group
equations for $U$ and $K$ take the form

\begin{eqnarray}
U^{\prime } &=&2(E_1-E_3) \\
K^{\prime } &=&(E_1+E_3)/2
\end{eqnarray}

\begin{figure}[h]
\begin{center}
\hbox{
\epsfxsize=2.5in
\epsfysize=2.5in
\epsfbox{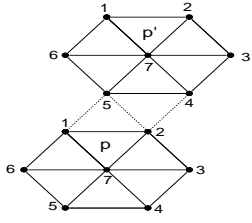}}
\end{center}
\caption{
Schematic diagram of the triangular lattice with hexagonal blocks.
Only two neighboring blocks $p$ and $p^{\prime }$ are drawn here. The dotted
lines represent the interblock interactions and solid line intrablock ones.}
\end{figure}

To illustrate this procedure, let us consider the triangular
lattice with hexagonal blocks as shown in figure 1.
For this non-bipartite lattice the interaction between blocks
can be written as

\begin{eqnarray}
V_{pp^{\prime }} &=&(-t)\sum_{\sigma ,i_1,i_1^{\prime } \newline
i_2,i_2^{\prime }} \{[|\Phi _{pi_1}><\Phi _{pi_1}|c_{1^{(p)}\sigma
}^{+}|\Phi _{pi_1^{\prime }}><\Phi _{pi_1^{\prime }}|]\times [|\Phi
_{p^{\prime }i_2}><\Phi _{p^{\prime }i_2}|c_{5^{(p^{\prime })}\sigma }|\Phi
_{p^{\prime }i_2^{\prime }}><\Phi _{p^{\prime }i_2^{\prime }}|]  \nonumber \\
+[|\Phi _{pi_1} &>&<\Phi _{pi_1}|c_{2^{(p)}\sigma }^{+}|\Phi
_{pi_1^{\prime }}><\Phi _{pi_1^{\prime }}|]\times [|\Phi _{p^{\prime
}i_2}><\Phi _{p^{\prime }i_2}|c_{4^{(p^{\prime })}\sigma }|\Phi _{p^{\prime
}i_2^{\prime }}><\Phi _{p^{\prime }i_2^{\prime }}|]  \nonumber \\
+[|\Phi _{pi_1} &>&<\Phi _{pi_1}|c_{2^{(p)}\sigma }^{+}|\Phi _{pi_1^{\prime
}}><\Phi _{pi_1^{\prime }}|]\times [|\Phi _{p^{\prime }i_2}><\Phi
_{p^{\prime }i_2}|c_{5^{(p^{\prime })}\sigma }|\Phi _{p^{\prime }i_2^{\prime
}}><\Phi _{p^{\prime }i_2^{\prime }}|]+H.C.\}
\end{eqnarray}

Since we would like to keep $V_{pp^{\prime }}$ of the form

\begin{equation}
V_{pp^{\prime }}=(-t^{\prime })\sum_\sigma [c_{p\sigma }^{\prime
+}c_{p^{\prime }\sigma }^{\prime }+H.C],
\end{equation}

we use the product transformation in Table I to simplify Eq. (20),

\begin{eqnarray}
V_{pp^{\prime }} &=&\sum_{\sigma ,<i,j>}\{<\sigma |_p^{\prime
}c_{i^{(p)}\sigma }^{+}|0>_p^{\prime }+[<-\sigma ,\sigma |_p^{\prime
}c_{i^{(p)}\sigma }^{+}|-\sigma >_p^{\prime }-<\sigma |_p^{\prime
}c_{i^{(p)}\sigma }^{+}|0>_p^{\prime }]n_{p-\sigma }^{\prime }\}c_{p\sigma
}^{\prime +}  \nonumber \\
\times \{ &<&0|_{p^{\prime }}^{\prime }c_{j^{(p^{\prime })}\sigma }|\sigma
>_{p^{\prime }}^{\prime }+[<-\sigma |_{p^{\prime }}^{\prime
}c_{j^{(p^{\prime })}\sigma }|\sigma ,-\sigma \downarrow \uparrow
>_{p^{\prime }}^{\prime }-<\sigma |_{p^{\prime }}^{\prime }c_{j^{(p^{\prime
})}\sigma }|0>_{p^{\prime }}^{\prime }]n_{p^{\prime }-\sigma }^{\prime
}\}c_{p^{\prime }\sigma }^{\prime }+H.C.,  \nonumber \\
( &<&ij>=<1,5>,<2,4>,<2,5>).
\end{eqnarray}
It can be easily seen now that in order to make all the extra couplings
vanish, it is necessary to make further restrictions upon the selected
states,

\begin{eqnarray}
&<&-\sigma ,\sigma |_p^{\prime }c_{i^{(p)}\sigma }^{+}|_p-\sigma >_p^{\prime
}=<\sigma |_p^{\prime }c_{i^{(p)}\sigma }^{+}|0>_p^{\prime }, \\
&<&-\sigma |_{p^{\prime }}^{\prime }c_{j^{(p^{\prime })}\sigma
}|\sigma ,-\sigma >_{p^{\prime }}^{\prime }=<0|_{p^{\prime }}^{\prime
}c_{j^{(p^{\prime })}\sigma }|\sigma >_{p^{\prime }}^{\prime }
\end{eqnarray}
Using similar calculations to the other neighboring interactions of the
block, we can finally obtain the following conditions,

\begin{equation}
<-\sigma ,\sigma |_p^{\prime }c_{i^{(p)}\sigma }^{+}|_p-\sigma >_p^{\prime
}=<\sigma |_p^{\prime }c_{i^{(p)}\sigma }^{+}|0>_p^{\prime }=\lambda
\end{equation}
for all the border sites on the block. Then the new hopping term becomes,

\begin{equation}
V_{pp^{\prime }}=\nu \lambda ^2\sum_\sigma c_{p\sigma }^{\prime
+}c_{p^{\prime }\sigma }^{\prime },
\end{equation}
where $\nu $ represents the number of couplings between neighboring blocks.
In figure 1 $\nu =3$.  The last renormalization group equation is 
readily obtained

\begin{equation}
t^{\prime }=\nu \lambda ^2t.
\end{equation}

Up to now, we have given a general discussion of the conditions
under which no proliferation of couplings results from the 
application of the BRG method to
non-partite lattice. Because on the border of non-partite lattice block,
there is only one type of sites, the above procedures can be extended to
other lattices with different dimensions or blocks without much difficulty.

\section{State Selections}

After deriving the conditions for the renormalization group equations,
the next task is to select states that satisfy these conditions.
At this stage the symmetry properties of the lattice  play an
important role. From Eqs.(23) and (24), it can be easily seen that if we assume
the particle number in the state $|0>^{\prime }$ to be $N_e-1$, then in
$\left| \uparrow >^{\prime },\right| \downarrow >^{\prime }$ and
$|\uparrow \downarrow >^{\prime },$ there should be $N_e,$ $N_e,$ $N_e+1$
particles respectively. Moreover if the spin in $|0>^{\prime }$ is $S_z$%
, the spins for $\left| \uparrow >^{\prime },\right| \downarrow
>^{\prime }$ and $|\uparrow \downarrow >^{\prime },$ should be
$S_z+1/2,S_z-1/2$ and $S_z$. The total electron number $N_e$ and the spin $S_z$ for each
block are good quantum numbers since their corresponding operators commute with the 
Hubbard Hamiltonian. So when we diagonalize the Hubbard Hamiltonian of the selected block, we keep 
$N_e$ and $S_z$ fixed to be ($N_e$-1, $S_z$), ($N_e$, $S_z$+1/2),  
($N_e$, $S_z$-1/2) and ($N_e$+1, $S_z$)
respectively. Thus we obtain four groups of eigenenergies and eigenstates 
corresponding to the above quantum numbers. From 
each group, we select the lowest-energy state to form the final
required four states. It should be mentioned that the lowest-energy state
has to be selected according to definite special symmetry considerations, 
which shall be discussed in the next paragraph. In order to obtain the 
insulating to conducting gap, which is defined to be the energy difference between extracting one 
electron from the system and adding one electron to it, 
$N_e$ is selected to be equal to $N_s$. For $S_z$, we choose it 
to be zero so as to make the block have the same spin property as the one-site. 
So now the
renormalized lattice will be composed of $N/n_s$ renormalized "sites"
with $N/n_s$ "particles".

Instead of forcing the above conserved
quantities upon the selected states in analogy to the one site properties in
a consistent way, here we get them directly from the
no-coupling-proliferation conditions. 
$\lambda $ does not depend on  $\sigma$ in Eq. (25), this
can be guaranteed by the particle-hole symmetry, which means that
only in half-filled lattice can the renormalized Hamiltonian have 
exactly the same form as the original one\cite{19}.
Moreover, the irrelevance of $\lambda$
with respect to the border site $i^{(p,b)}$ can be shown by requiring the
selected states to belong to the same irreducible representation of the
spatial group of the lattice. For the triangular lattice with hexagonal blocks,
the Hamiltonian is invariant under $C_{6v}$ \cite{20}. So if we choose the
same one-dimensional irreducible representation of the group $C_{6v}$ for
$|0>^{\prime }$ , $\left| \uparrow >^{\prime },\right| \downarrow
>^{\prime }$ and $|\uparrow \downarrow >^{\prime },$ the conditions can be
satisfied.

\section{Numerical Calculations and discussions}

The insulating to conducting gap $\Delta_g $\cite{3} and the ground
state energy per site $E_g$  for the half-filled
Hubbard model on a triangular lattice with hexagonal blocks can be written as

\begin{equation}
\Delta_g = {lim}_{n\rightarrow \infty } \;\; U^{(n)}
\end{equation}

and

\begin{equation}
E_g = {lim}_{n\rightarrow \infty } \;\; \frac{K^{(n)}+\frac{U^{(n)}}4}
{7^n}+\frac U2
\end{equation}

\noindent where $n$ denotes the number of iterations in the renormalization equations.
As ${n\rightarrow \infty}$, the lattice containing ${7^n}$ sites will be renormalized
to one 'site', which can be solved exactly to obatin the above equations. U/2 is
added in Eq.(29) in order to compensate the chemical potential energy substracted
in the original Hamiltonian.
 The metal-insulator-transition for
  this lattice can be examined by considering  the energy gap $\Delta 
$ as a function
of $(U/t)$. If the insulating gap disappears the system
exhibits metallic  behavior. Otherwise it is insulating.

\begin{center}
\begin{figure}[h]
\hbox{
\epsfxsize=2.5in
\epsfysize=2.5in
\epsfbox{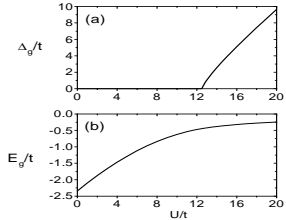}}
\caption{
$U/t$ dependence of (a) the insulating gap and (b) the ground-state
energy per site. }
\end{figure}
\end{center}

Figure 2 shows the numerical results
for the scaled gap $\Delta_g/t$  as a function of $(U/t)$. There is a
first-order phase transition at $(U/t)_c\approx 12.5$.
This finding is quite different from the case
of half-filled square lattice where an insulating gap
exists for arbitrarily small values of $U/t$.
This is because there is a
perfect nesting of the Fermi surface on a square lattice, which makes the
model unstable toward antiferromagnetism as soon as a nonzero $U$ is turned
on, driving the system to an insulating state.

On the triangular lattice, for the lack of perfect nesting,
metal-insulator-transition takes place at a
finite value of $U/t$.  But because there is no exact solution to the
two-dimensional Hubbard model, there are still controversies upon the exact
value for this critical point. The situation becomes more complicated and
subtle because of the frustrations inherent in the triangular lattice, which
may induce a nontrivial competitions among different magnetic phases. In the
Hartree-Fock calculations by Krishnamurthy et al.\cite{21,22} they
found that for small $U/t$ the system is a paramagnetic metal which
turns to a metal with incommensurate spiral spin-density wave
at $U_{c1}/t=3.97.$ Two successive first-order phase transitions occur
at $U_{c2}/t=4.45,$ a semi-metallic linear
spin-density wave is stabilized and a first-order 
metal-insulator-transition to an
anti-ferromagnetic insulator occurs at $U_{c3}/t=5.27.$ Capone
et al.\cite{23} obtained qualitatively similar phase transitions by
using  the Kotliar-Ruckenstein slave-boson technique.
That is, the
weak-coupling paramagnetic metal continuously evolves into a spiral metal at
$U_{c1}/t=6.68,$ which crosses the linearly polarized spin-density-wave
ground state at $U_{c2}/t=6.84.$ The latter phase  undergo a further
first-order transition toward an anti-ferromagnetic insulator at
$U_{c3}/t=7.68$. The exact diagonalization results exhibit
a first-order transition between the paramagnetic metal and the
anti-ferromagnetic insulator at $U/t=12.07,$ without intermediate
''exotic'' phases\cite{24}.
Our  results for the critical value of $(U/t) \approx 12.5$ for the 
metal-insulator-transition
and the ground state energy as a
function of $U/t$ are  in agreement with results obtained by
the exact diagonalization method\cite{24}.

In summary, we  find that there
is a first-order phase  metal-insulator-transition,
at $(U/t)_c=12.5$ in full agreement with the exact diagonalization 
result $U/t=12.07$.

\section{Concluding Remarks}

A practical renormalization method for an array described by a 
Hubbard Hamiltonian has been presented. Specifically, we could find 
conditions under which the renormalized Hubbard Hamiltonian on a 
nonpartite lattice has no proliferation of coupling. The method is 
especially suited for getting the lowest excited states. An 
application demonstrating the identification of a first order 
metal-insulator transition has been discussed in detail.
In future work we intend to build upon the present results in two 
directions. First, to apply them to the computation of the transport 
properties of lattices of quantum dots. We also hope to be able to 
incorporate into the method one essential aspect of quantum dots 
that, so far, is not included. This is the (small but finite) 
variability of the chemical potential. This results from the size 
fluctuations of dots as they are prepared in the laboratory.
 In real-space BRG method, how to truncate the states for one block is of central
importance to the accuracy of the final results. In this paper, four states for each 
block are selected from the requirement of non-proliferations of parameters. This
greatly simply the theoretical handling of this problem while keeping the main related 
physics. But for two-dimensional system,  how to judge the convergence when more states are kept is still an 
open problem in this direction. More work is under way.

\vspace{1cm}
\noindent {\bf \large Acknowledgements} 
\vspace{0.2cm}
\newline \normalsize \newline We are delighted to dedicate this paper to Professor R.S. Berry on the
occasion of his 70th birthday. We acknowledge the financial support of the
National Science Foundation and the Office of Naval Research.
 
 % BIBLIOGRAPHY
 
\newpage
\begin{figure}[tbp]

\caption{$U/t$ dependence of (a) the insulating gap and (b) the ground-state
energy per site. }

\end{figure}


\newpage
\begin{thebibliography}{70}

\bibitem{r1} Markovich, G; Collier, C. P.; Henrichs, S. E.; Remacle, F.; Levine, R. D.; Heath,
J. R.; Accounts of Chemical Research 1999, 32, 415-423.

\bibitem{r2} Remacle, F.; Levine, R. D., Proc. Natl. Acad. Sci. USA 2000, 97, 553. Remacle, F.; Levine, R. D.,
ChemPhysChem. 2001, 2, 20

\bibitem{1}
Hubbard,J., Proc. R. Soc. Longdon A 1963, 276, 238; A 1964, 277, 237;A 1964, 281, 401.


\bibitem{r3} Remacle F.; Beverly,K.C.; Heath, J. R.; Levine,R. D., J. Phys. Chem. in press.

\bibitem{r4} Mott, N. F., Metal-Insulator Transitions; Taylor and Francis: London, 1990.



\bibitem{2}  Lieb,E. H., arXiv: cond-mat/9311033, 1993; Tasaki,H. arXiv: cond-mat/9512169, 1997; arXiv: cond-mat/9712219, 1997; and references therein.

\bibitem{3} Lieb E.; Wu, F., Phys. Rev. Lett. 1968, 20, 1445.

\bibitem{4} Senechal,D.; Perez, D.; Pioro-Ladriere,M., Phys. Rev. Lett. 2000, 84, 522.

\bibitem{5} White, S. R.; Scalapino, D. J.; Sugar, R. L.; Loh, E. Y. Jr.; Gubernatis, J. E.; Scalettar, R. T., Phys. Rev. B 1989, 40, 50.

\bibitem{6} Penn, D., Phys. Rev. 1966, 142, 350.

\bibitem{7} Tahir-Kheli, R.; Jarrett, H., Phys. Rev. 1968, 180, 544; Roth, L., Phys. Lett. 1981, 20, 1431; Kishore, R.; Joshi, S., Phys. Rev. 1969, 186, 484; Arai, T.; Parinellok, M., Phys. Rev. Lett. 1971, 27, 1226.

\bibitem{8} Cyrot, M., J. Phys. (Paris) 1972, 33, 125; Kimball, J.;  Schriieffer, J. R., Int. Conf. on Magnetism, Chicago(1971); Cyrot, M., Phys. Rev. Lett.1970, 25, 871; Phil. Mag. 1971, 1031; Cyrot, M.; Lacour-Gayet, P., J. Phys. C 1974, 7, 400.

\bibitem{9} Gutzwiller, M. C., Phys.Rev. 1965, 137, A1726; Phys. Rev. Lett. 1963, 10, 159; Langer, W.; Plischke, M.; Mattis, D., Phys. Rev. Lett. 1969, 23, 149; Kaplan, T.; Bari, R., J. Appl. Phys. 1970, 41, 875; Brinkman, W.; Rice, T. M., Phys. Rev.B 1970, 2, 4302; Chao,  . A., Phys. Rev.B 1971, 4, 4034.

\bibitem{10}  Harris, A. B.; Lange, R. V., Phys. Rev. 1967, 157,295; Hubbard, J., Proc. Roy, Soc(London) A 1966, 296, 100; Kanamori, J., Progr. Theor. Phys. (Kyoto) 1963, 30, 275; Esterling, D.; Lange, R. V., Rev. Mod. Phys. 1968, 40, 796.

\bibitem{11}  Hirsch, J. E., Phys. Rev. B 1979, 20, 3907, Phys. Rev. B 1980, 22, 5259.

\bibitem{12}  White, S. R., Phys. Rev. Lett. 1992, 69, 2863; Phys. Rev. B 1993, 48, 10345.

\bibitem{12a} Tokuyasu, T.; Kamal, M.; Murthy, G., Phys. Rev. Lett. 1993, 71, 4202.


\bibitem{12b} Murthy, G; Kais, S., Chem. Phys. Lett. 1998, 290, 199.


\bibitem{13} Dasgupta, C.; Pfeuty, P., J. Phys. C 1981, 14, 717; Jullien, R.; Pfeuty, R.; Fields, J. N.; Doniach, S., Phys. Rev. B 1978, 18, 3568.

\bibitem{14} Perez-conde, J.; Pfeuty, P., Phys. Rev. B 1993, 47, 856; Vanderzande, C., J. Phys. A: Math. Gen. 1985, 18, 889.

\bibitem{15} Bhattacharyya, B.; Sil, S., J. Phys.:Condens. Matter 1999, 11, 3513.

\bibitem{16} Topics in current physics, {\it Real-space renormalization}, Editor: T. W. Burkhardt and J. M. J. van Leeuwen, Springer-Verlag Berlin Heidelberg New York, 1982.

\bibitem{17}  Imada, M.; Fujimori, A.; Tokura, Y., Rev. Mod. Phys. 1998, 70, 1039.

\bibitem{19} Bhattacharyya, B.; Sil, S., Phys. Lett. A 1993, 180, 299.

\bibitem{20}  Boardman, A. D.; O$^{\prime }$Connor, D. E.; Young, P. A., {\it Symmetry and its application in science}, John Wiley and Sons, New York, 1973.

\bibitem{21}  Krishnamurthy, H. R.; Jayaprakash, C.; Sarker, S.; Senzel, W., Phys. Rev. Lett. 1990, 64, 950.

\bibitem{22}  Jayaprakash, C.; Krishnamurthy, H. R.;  Sarker, S.; Wenzel, W., Europhys. Lett. 1991, 15, 625.

\bibitem{23}  Capone, M.; Capriotti, L.; Becca, F.; Caprara, S., Phys. Rev. 2001, B63, 0854104.

\bibitem{24}  Brinkman, W. F.; Rice, T. M.,  Phys. Rev. B 1970, 2, 4302.

\end{thebibliography}
\end{document}